\documentclass{egpubl}
\usepackage{eurovis2021}

\SpecialIssuePaper         

\usepackage[T1]{fontenc}
\usepackage{dfadobe} 
\usepackage{amssymb}
\usepackage{amsmath}

\usepackage{cite}  
\BibtexOrBiblatex
\electronicVersion
\PrintedOrElectronic
\ifpdf \usepackage[pdftex]{graphicx} \pdfcompresslevel=9
\else \usepackage[dvips]{graphicx} \fi

\usepackage{egweblnk}
\usepackage{xcolor}
\definecolor{revision}{RGB}{0,136,55}

\title[Visualizing Carotid Blood Flow Simulations]%
      {Visualizing Carotid Blood Flow Simulations for Stroke Prevention}

\author[P. Eulzer et. al.]
{\parbox{\textwidth}{\centering P. Eulzer$^{1}$, M. Meuschke$^{1,2}$, C. M. Klingner$^{3}$ and K. Lawonn$^{1}$
        }
        \\
{\parbox{\textwidth}{\centering $^1$University of Jena, Faculty of Mathematics and Computer Science, Germany\\
         $^2$ University of Magdeburg, Department of Simulation and Graphics, Germany \\
         $^3$ University Hospital Jena, Clinic for Neurology, Germany
       } 
}
}

\begin{document}


\maketitle
\begin{abstract}
In this work, we investigate how concepts from medical flow visualization can be applied to enhance stroke prevention diagnostics.
Our focus lies on carotid stenoses, i.e., local narrowings of the major brain-supplying arteries, which are a frequent cause of stroke.
Carotid surgery can reduce the stroke risk associated with stenoses, however, the procedure entails risks itself.
Therefore, a thorough assessment of each case is necessary.
In routine diagnostics, the morphology and hemodynamics of an afflicted vessel are separately analyzed using angiography and sonography, respectively.
Blood flow simulations based on computational fluid dynamics could enable the visual integration of hemodynamic and morphological information and provide a higher resolution on relevant parameters.
We identify and abstract the tasks involved in the assessment of stenoses and investigate how clinicians could derive relevant insights from carotid blood flow simulations.
We adapt and refine a combination of techniques to facilitate this purpose, integrating spatiotemporal navigation, dimensional reduction, and contextual embedding.
We evaluated and discussed our approach with an interdisciplinary group of medical practitioners, fluid simulation and flow visualization researchers.
Our initial findings indicate that visualization techniques could promote usage of carotid blood flow simulations in practice.\\
%


\end{abstract}
\section{Introduction}
Stroke is the second most frequent cause of death worldwide~\cite{burden_of_stroke}.
In non-fatal cases, stroke often leads to permanent neurological disabilities and post-stroke care is associated with substantial economic costs~\cite{rajsic2019economic}.
The increasing burden of stroke, which is partly attributed to the demographic shift towards an older population, calls for more effective prevention strategies~\cite{gorelick2019global}.

The majority of strokes, about 87\%, are caused by limited blood supply to the brain~\cite{donnan_stroke}.
They typically arise from a localized narrowing of arteries (a \textit{stenosis}), caused, for example, by atherosclerosis.
Stenoses occur most frequently in the carotids and their successive branches.
The carotids are two large arteries responsible for the majority of cranial blood supply.
A possible preventive option is an early treatment of developing stenoses through stent insertion or the surgical removal of plaque, thereby averting full closure.
This also alleviates the possibility of the separation of plaque particles, which can be carried upstream and often block smaller vessels.
Carotid surgery, however, entails its own risks, which need to be carefully weighed against the probability of stroke~\cite{halm2009risk}.
It is, therefore, crucial to choose the correct moment for surgical intervention and the optimal treatment strategy.
Today, two concurrent methods are widely implemented in routine diagnostics: Computed tomography angiography (CTA) and duplex ultrasonography.
CTA is used to determine vessel \textit{morphology}, i.e., the 3D shape.
Sonography reveals some \textit{hemodynamic} properties, describing the blood flow.
The parameters measured with each modality need to be mentally combined to derive a decision.

Recent years have seen advances in simulated hemodynamics, which derive properties of the flow field based on computational fluid dynamics (CFD)~\cite{caballero2013review, SZAJER201862, saqr2020does}.
This emerging domain could substitute or complement the information gained through sonographic assessments.
CFD can generate high-resolution hemodynamic information based on patient-specific morphology and can also be applied to regions where ultrasound signals are distorted or occluded.

The comprehensive work covering medical flow visualization shows that carefully designed exploration methods can support the transfer of simulated hemodynamics into medical research and practice~\cite{oeltze2019generation}.
We propose to extend these efforts to cover the domain of stenosis analysis in the carotids.
In particular, better visualization tools would allow the integration of hemodynamic and morphological features, which are currently separately assessed.
In this paper, we outline our reasoning for how medical practitioners could navigate the spatio-temporal data space faster and more effectively by using sophisticated exploration methods.
We developed an application that integrates these methods, attempting to aid in the identification, comparison, and evaluation of stenoses.
Ultimately, our goal is to facilitate clinical decision making and, therefore, advance stroke prevention.

As this work is motivated by our conjecture that visualizing blood flow in the carotids could support real-world analysis tasks, we follow the recently proposed outline to a \textit{data-first} design study~\cite{oppermann2020data}:
We contribute a stakeholder analysis, concluding that medical practitioners could benefit from simulated flow in the human carotids.
We then analyze and abstract the data and tasks involved in studying blood flow properties w.r.t. stroke prevention.
Based on our conclusions, we distill the most promising methods developed in medical flow visualization, which we adapt for the exploration of carotid hemodynamics and morphology.
We implemented and tested these methods and discussed our approaches with an interdisciplinary group of physicians, simulation researchers, and flow visualization experts.
\section{Background}
Our work is positioned at the intersection of stroke prevention diagnostics and flow simulation. Therefore, we will now summarize context and terminology from the medical and CFD domains.

\subsection{Medical Background}
\begin{figure}[tbh]
   \centering
   \includegraphics[width=1.0\linewidth]{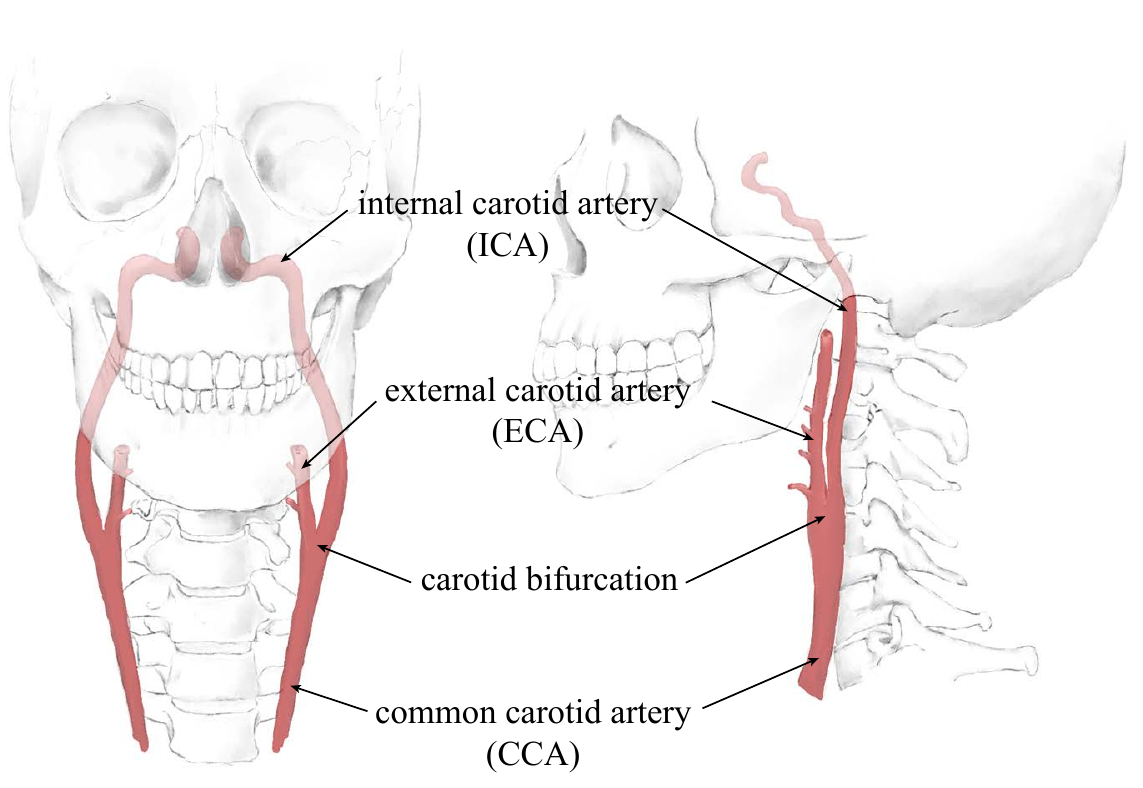}
   \caption{\label{fig:anatomy}
     Anatomy of the carotids.}
\end{figure}

Stroke-causing stenoses frequently affect the carotids and following branches.
Timely surgical intervention can prevent stroke as a result of carotid stenosis~\cite{halliday201010, Orrapin99}, which makes the two arteries prime candidates for preventive diagnostics and treatments.
Each carotid consists of two branches, the internal and external carotid artery (ICA and ECA), as shown in Figure~\ref{fig:anatomy}.
They emerge from the common carotid artery (CCA).
The point of separation is referred to as carotid bifurcation.
It is a predilection site for the buildup of atherosclerotic plaque and stenosis formation.

Patients who suffered a stroke or show symptoms that may indicate possible stenoses are typically subjected to sonography of the carotids.
Multiple parameters are measured, which mostly relate to the hemodynamics of the blood flow.
The constricted vessel walls of a stenosis generally lead to an increase in flow speed, which abruptly drops behind the stenosis.
This sudden reduction in velocity leads to turbulence that also becomes visible.
Duplex sonography is a popular method as it is fast, cheap, and non-invasive.
However, it is difficult to assess intracranial regions with ultrasound, as they are occluded by the skull.

In many cases, an angiography is also necessary, often in form of a CTA, to reveal the intracranial vessels and give a more accurate representation of the morphology.
The injection of a contrast agent highlights the vessel \textit{lumen}, i.e., the regions where blood actively flows, in the resulting volume images.
This allows determining the position and size of a stenosis, based on which its severity can be derived.
However, CTA and similar routinely used methods reveal no temporal information about the flow.

\subsection{Computational Hemodynamics}
CFD describes the simulation of a fluid’s motion and its interaction with other elements.
On the macroscopic level blood is a fluid, i.e., it can be characterized by properties like density and viscosity. 
Therefore, the hemodynamics (the dynamics of blood flow) can be derived from CFD simulations.

Typically, the problem space is defined by vessel walls, which are either artificially modeled or segmented from volume images.
The space is partitioned into small cells containing descriptions of the flow properties, such as velocity and pressure.
Flow may enter the domain at predefined inlets and exits at designated outlets.
Given initial conditions for all variables as well as input or boundary changes that may occur over time, the internal fields can be solved.
If physical time is required, the discretized domain can be iteratively updated over temporal increments.
Often, a pulsatile flow is defined at the inlets, which models the cardiac cycle based on population averages.
\section{Related Work}
Visualizations can facilitate CFD research, by making result interpretation possible.
Furthermore, they provide a means of transfer into medical practice, where clinicians can benefit from depictions tailored to specific data exploration tasks.
Specialized techniques have been proposed for three areas in particular: hemodynamics of aneurysms, blood flow in the aorta, and nasal airflow, as outlined in the state-of-the-art report by Oeltze-Jafra et al.~\cite{oeltze2019generation}.
To the best of our knowledge, flow analysis w.r.t. carotid stenoses has not yet been covered from a visualization standpoint.
Therefore, this section refers to aspects of related works in the domain of blood flow visualization, from which methods and findings can be generalized and adapted to match our specific requirements.
For a more detailed introduction into the challenges and techniques in biomedical flow analysis, we refer to the survey by Vilanova et al.~\cite{vilanova2014visual}.

\textbf{Overview.} Generally, users first require an overview of the properties across the spatio-temporal domain.
These can refer to the hemodynamics, wall-related attributes, or both~\cite{meuschke2016combined}.
A useful overview depiction guides towards points of interest.
Wall-related parameters can be shown using colormaps, textures, and opacity~\cite{appanaboyina2009simulation}.
The visualization of 3D flow fields generally relies on either integral curves, arrow glyphs, or particles~\cite{buonocore1998visualizing}.
More advanced approaches take these ideas further and perform, e.g., clustering of flow curves~\cite{meuschke2017glyph, oeltze2014blood, van2012visualization}.
Information on vessel walls can be combined with internal flow field data through dynamic cutaway views, which show both surface and flow-related parameters~\cite{lawonn2015occlusion}.
Straightened representations, for instance of the aorta~\cite{angelelli2011straightening}, result in simplified depictions that allow at-a-glance assessments~\cite{kohler2013semi, oeltze2014blood}.
Tree-like structures, such as the vasculature, can be transferred to graphs or maps that display the underlying connectivity~\cite{al2014neurolines, borkin2011evaluation, lichtenberg2020parameterization}.
In general, dimensional reduction techniques have shown great effect when creating overview visualizations in the medical domain~\cite{kreiser2018survey}.

\textbf{Temporal analysis.} Biomedical flow is intrinsically time-dependent.
This aspect of the data often needs to be shown, e.g., to allow investigation of stability or to assess the flow behavior over the cardiac cycle~\cite{meuschke2018exploration}.
The general goal is to display the temporal evolution of the data and to highlight important spatio-temporal features.
Typically, consecutive time-steps of a simulation are shown in a juxtaposition~\cite{feliciani2015multiscale, wong2009cardiac}.
However, the interpretation of such depictions can be time-intensive and exhausting.
Therefore, advanced approaches facilitate animation or integrated displays of multiple time-steps.
For example, integral lines can be highlighted~\cite{van2010exploration} or animated~\cite{lawonn2014adaptive} according to the point in time.

\textbf{Probing.} After familiarizing themselves with the domain, users often need to further explore specific regions of interest.
For this purpose, many visualization tools offer the ability to probe flow fields at specified locations, resulting in the output of numerical values or more detailed views.
Common approaches are cut planes, which show 2D cross-sections of the 3D domain~\cite{bates2015dynamics}.
More advanced ideas include flexible probing geometries that can be arbitrarily placed or dragged through the 3D object.
Such \textit{slice widgets} can, for instance, show the contour of walls at the cross-section of vessels~\cite{glasser2014combined, markl2010coregistration}.
Similarly, the use of a \textit{flow lense} was proposed to separate the domain into a focus (inside) and context region (outside)~\cite{gasteiger2011flowlens}.

\textbf{Contextualization.} Context rendering of the anatomy surrounding a focused structure, like a vascular tree, can support spatial orientation and correlation of flow and morphology~\cite{oeltze2019generation}.
The objective of the underlying task is to understand the initiation and progression of possible pathology.
A typical approach to display context information is to use semi-transparent surfaces, e.g., of the near vasculature~\cite{andersson2015quantitative, gambaruto2012flow}.
For example, cerebral aneurysms can be embedded within the arterial system~\cite{hastreiter1998fast}.
Advanced visualization methods are often specific regarding either general structures (far context) or directly connected tissue (near context)~\cite{oeltze2019generation}.
Far context might refer to bones, vessels, or tissue that could help to understand the location and orientation of the focus structure~\cite{higuera2003enhanced, higuera2004automatic}.
%
%
%
Near context is often associated with vessel walls, a display of which is useful to provide context for the internal flow field~\cite{gasteiger2010adapted}.
%
%

\begin{figure*}[tbh]
    \centering
    \includegraphics[width=0.9\linewidth]{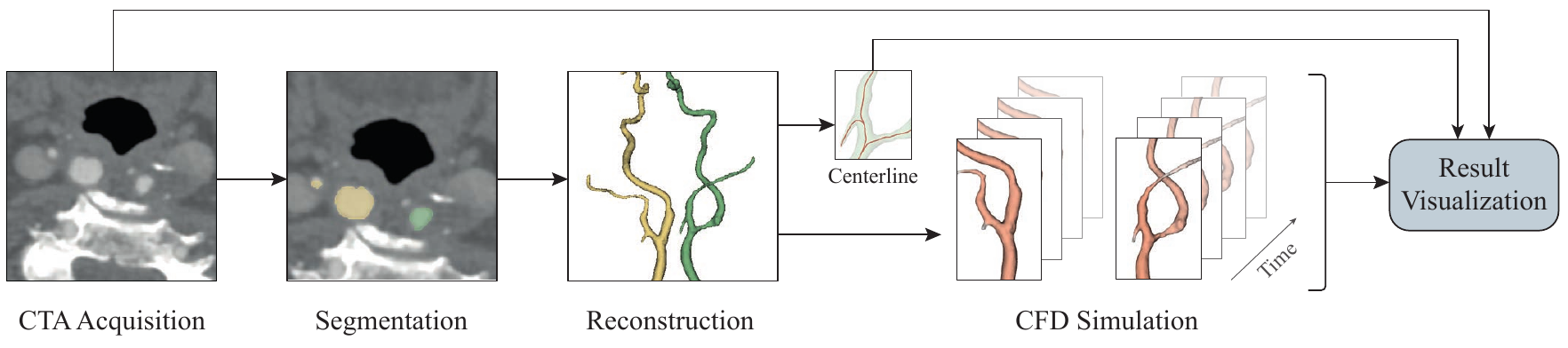}
    \caption{The pipeline used in this work. The carotid lumen is reconstructed from CTA volumes. Based on the models, a simulation of the hemodynamics is performed, resulting in discrete flow fields for both sides over multiple time-steps. The simulation results can then be explored with our framework. Navigation relies on the centerline of the segmented models and we incorporate the original data for contextualization and validation purposes.}
    \label{fig:pipeline}
\end{figure*}
\section{Domain Characterization}
The vascular models used in blood flow simulations are often segmented from already available volume images, which means that no additional diagnostic methods are necessary to acquire the data.
In the case of carotid hemodynamics, CTA scans are routinely performed, vessel segmentation algorithms based on these scans exist~\cite{MANNIESING2010759}, and a variety of CFD techniques were developed~\cite{LOPES2020110019}.
On this basis, we analyzed who would profit from this data, which types of tasks exist, and whether visualization methods needed to be adapted to better facilitate these use cases.
We took this data-first perspective~\cite{oppermann2020data} to better understand the challenges and opportunities the data poses before applying it to a problem.
Further, we intend to provide an initial frame to the domain of visualizing flow in the carotids and outline possible directions.

\subsection{Data Abstraction}
CFD simulations of blood flow span a wide range of methods producing different results.
On an abstract layer, however, the types and semantics of the outputs are comparable.
The simulation computes attribute fields over the domain, which can include scalars (e.g., pressure) and vectors (e.g., velocity).
Each field is resolved in a spatial dimension, which can either refer to the volume inside of the vessel or the wall.
Spatial data granularities contain the \textit{region} of the vascular tree used in the simulation and the \textit{spatial resolution}, describing the number and size of individual cells.
If physical time is simulated, the fields are also temporally resolved.
Temporal data granularities are the \textit{aggregate}, for example, the number of cardiac cycles that are simulated in total, as well as the \textit{temporal resolution}, which effectively describes the number of discrete time-steps that are output.

In general, CFD allows full control over the spatial and temporal resolutions and the aggregate.
The simulated region is only constrained by the lumen visibility in the underlying volume data.
This creates opportunities for flow analysis beyond the capabilities of traditional diagnostic methods, as hemodynamic and morphological information is intrinsically combined in the simulation domain.
In addition to values like velocity that are already used in clinical routine, simulations can compute supplementary parameters which could be linked to the pathophysiology of stenoses, such as wall shear stress~\cite{younis2004hemodynamics}.
It should be noted, however, that computational flow models necessarily rely on simplifications, which poses challenges.
The role of CFD to derive hemodynamic factors is controversial due to the difficult validation of the generated data~\cite{xiang2014cfd}.
Results have been shown to vary~\cite{ScheidRehder2019VCBM}, depending on the employed methods of segmentation, mesh creation, and flow simulation.
Only gradually, increasingly sophisticated methods are able to produce robust calculations that are reasonably close to measured parameters~\cite{SZAJER201862}.

\subsection{Stakeholder Analysis}
Currently, medical and simulation researchers are actively working with flow data generated by CFD in various domains, including carotid hemodynamics.
Their tasks evolve around validation and comparison of models, as well as hypothesis forming and testing.
The second group of stakeholders who could profit from detailed carotid flow parameters are medical practitioners, foremost radiologists and neurologists.
The additional information could help make therapeutic decisions, plan interventions, and evaluate treatment success.

We reached out to several independent stakeholders: a flow simulation researcher (13 years of experience) working with carotid data sets, a neurologist (15 years of practice), and a radiologist (30 years of practice).
Both physicians are involved in the treatment of stroke patients.
We discussed the prospects of simulated blood flow analysis for stroke prevention, the current state-of-the-art in each field, and whether adapted visualizations could facilitate tasks by providing insight into the data.

Medical and simulation researchers use workflows that rely on standard depictions of the result data, e.g., color-coded vessel wall geometry.
These depictions could presumably be improved to enable more direct comparisons and integrated evaluations of multiple parameters.
However, when speaking with medical practitioners, we found another issue:
Even though CFD use for carotid hemodynamics has been extensively studied~\cite{LOPES2020110019}, the transfer to medical practice is low.
A possible reason is that such simulations miss validations and derived implications seem to have a small impact~\cite{xiang2014cfd}.
Another possibility might be that the generated data is difficult to grasp for clinicians and does not target their immediate necessities.
Most likely, a combination of both factors is involved, which makes the applied use of computed carotid blood flow difficult.
If the data is not broadly validated it cannot be used to derive implications and will not be integrated into clinical workflows.
However, if it is not integrated, large-scale trials that could validate the data are less likely.
We believe that exploration and visualization methods tailored to clinical needs could help bridge the gap between research and application.
For this reason, we decided to focus on possible clinical use cases in this work.
We conjecture that visualization methods developed to derive insights from other types of hemodynamic simulations, like blood flow in aneurysms, can be selectively adapted to fit these scenarios.

\begin{figure*}[tbh]
    \centering
    \includegraphics[width=1.0\linewidth]{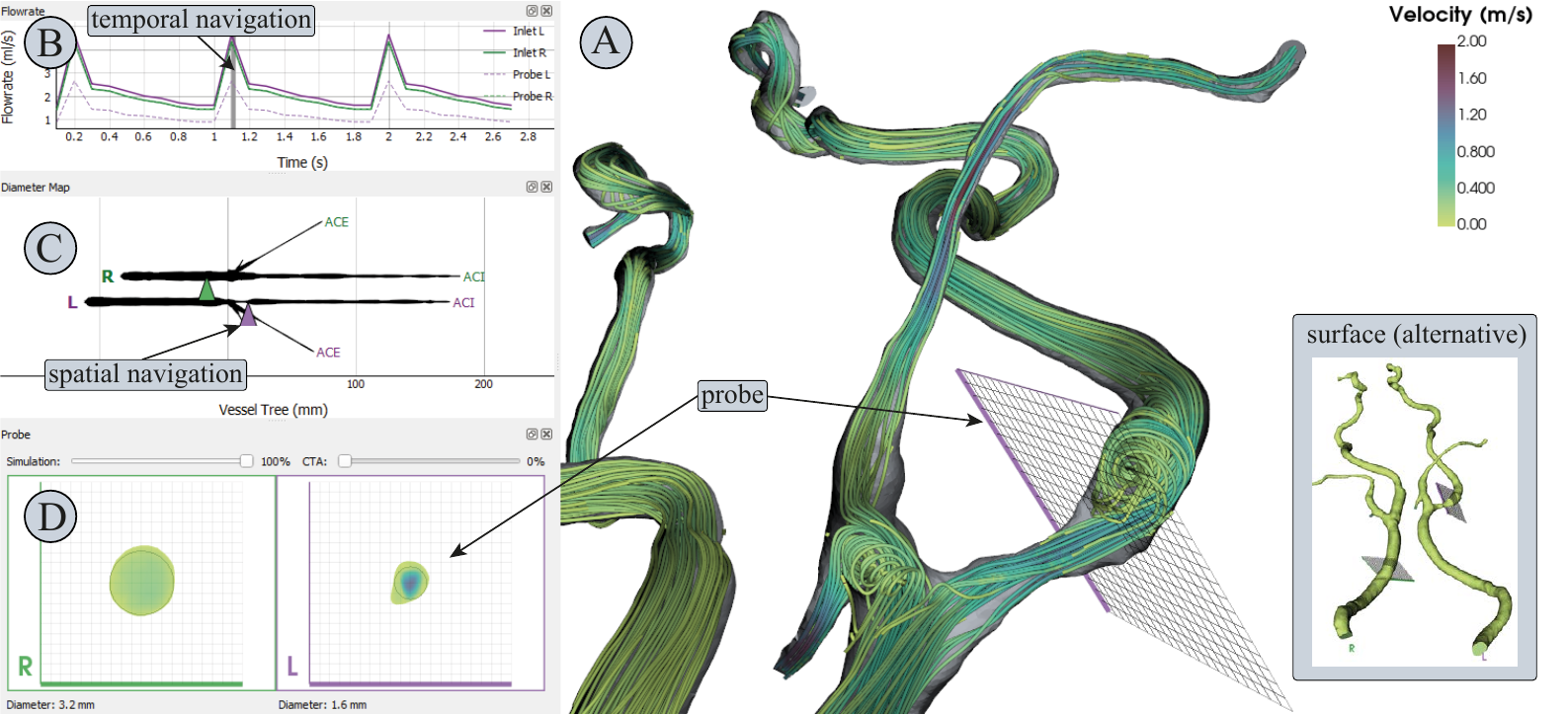}
    \caption{Annotated screenshot of the developed application. Four views are linked: a 3D depiction of the domain (A), a graph of the flow rate (B), a map of the carotids (C), and cross-sections of the vessels at the probe location (D). The 3D view either displays a surface model or streamlines.}
    \label{fig:overview}
\end{figure*}
\subsection{Task Abstraction}
Our goal is to identify which types of visual displays are required to map the information derived from CFD to the tasks involved in assessing carotid stenoses.
We iteratively discussed possible approaches with the physicians and found that the typical tasks performed when analyzing stenoses condense to the following:
First, possible stenoses need to be identified.
This is a substantial task, as often patients exhibit more than one critical region with a buildup of atherosclerotic plaque.
Second, morphological features are analyzed w.r.t. questions like:
\textit{What is the diameter inside and after the stenosis? Over which length is the vessel constricted?}
If precise values are needed, they are usually measured based on CTA.
We found that these first two aspects require large parts of the vessel to be visible, as the morphology of the whole artery is taken into account.
Third, hemodynamic features are analyzed, typically based on sonography.
Physicians are mostly interested in the two time-points of highest (systole) and lowest flow (diastole).
\textit{What is the peak velocity? Is turbulence visible?}

In addition to the primary tasks, further inquiries revealed processes that are performed in parallel and might not be obvious at a first glance but are equivalently important.
For one, the location of a stenosis determines possible treatment and intervention strategies.
An intracranial stenosis needs to be handled differently than one at the bifurcation.
This is a trivial factor, however, it needs consideration, as visualizations should not remove this context information.
Further, it should be factored in that full assessments include the analysis of tissue around a stenosis.
If plaque causes the vessel constriction, its type (e.g., soft or calcified) and configuration (e.g., even or fractured) can be used to judge the pathophysiological factors at play.
This answers questions like: \textit{How fast did the stenosis develop? How likely is the rupture of plaque?}
Last but not least, the left and right-side carotid are often contrasted for a full picture to inquire, for example: \textit{Does one side compensate for a stenosis in the other?}
We summarize these findings in four types of tasks that a visualization needs to cover:
\begin{description}
\itemsep0.5em
    \item[T1] Gain an overview of the attribute space and find relevant points.
    \item[T2] Probe selected spatio-temporal data points.
    \item[T3] Compare observations regarding time and side (left vs. right).
    \item[T4] Explore the context of global and local anatomical features.
\end{description}

We presume that CFD simulations of carotid blood flow could facilitate these tasks and enhance state-of-the-art diagnostics.
Routine CTAs could serve as a sensible basis for further data processing.
If a sufficiently broad domain is segmented \textbf{T1} could be assisted.
A 3D model reconstruction in combination with computed flow fields could further support advanced probing (\textbf{T2}).
Fine spatial resolutions can be achieved in the simulation and morphological attributes like the diameter are accessible.
A temporal resolution is also available.
For comparison (\textbf{T3}), flow in the left and right carotid can be computed and the results can be displayed in a synchronized visualization.
Lastly, the original CTA and the processed simulation data complement each other.
Therefore, we propose an integrated visualization, which incorporates the CTA volume and allows to evaluate near and far context (\textbf{T4}).

\section{Pre-Processing}
We obtained 13 CTA data sets of patients with differing degrees and locations of carotid stenoses.
From these volumes, we segmented the lumen of the full CCA and ICA on both sides until the vessels split into smaller branches.
We also include parts of the ECA close to the bifurcation.
The ECA is less important for diagnostic purposes but required for a correct flow simulation.
We used the level-set method from the VMTK~\cite{VMTK} and subsequent manual corrections to perform the segmentations.
Then, we computed the centerline of the extracted vessel tree based on the 3D Voronoi diagram of the geometry, as this method yields robust and accurate results~\cite{wang2010comparisons}.
Each centerline consists of around 3000 points, providing smooth and consistent lines.
Furthermore, the vessel radius is computed as a convenient byproduct, since the procedure aims to maximize the enclosed spheres.

We used the freely available toolkit OpenFOAM to generate volume meshes of the segmented carotid lumen and perform the CFD simulation.
We intended to produce physiologically plausible data that allows us to implement and evaluate visualization methods to test our assumptions.
Therefore, we used a standard velocity profile at the inlet of the CCA which models a typical flow waveform of the cardiac cycle~\cite{antiga2002patient}.
Solving the Navier-Stokes equations lead to patient-specific flow information, where blood is considered an incompressible Newtonian fluid.
For each data set, we simulated three cardiac cycles (about 2.7 seconds) from which we saved 30 equidistantly spaced time-steps.
The full pipeline is illustrated in Figure~\ref{fig:pipeline}.
\section{Methods}
We chose and adapted a variety of methods developed in flow visualization and implemented them in a framework which allows simultaneous exploration of the morphology and hemodynamics of both carotids.
Over our implementation cycles, we converged on the layout shown in Figure~\ref{fig:overview}, combining four linked views.
We present the 3D domain as either a surface rendering or a streamline visualization (A).
The data space navigation is split into the temporal and spatial components of the CFD results.
Orientation in the temporal dimension is facilitated by a line graph showing the volumetric flow rate in each vessel (B).
This depiction is representative of the simulated cardiac cycles.
A sought time-step can be selected by dragging a slider across the graph.
The spatial orientation is aided by a map-like flattening of the vessel tree (C), which provides at-a-glance information on the vessel diameter.
The location of the probes is determined by markers in this reduced depiction, which can be dragged over the domain and snap to the nearest vessel.
The probe views (D) show cross-sections of the vessels at the selected spatio-temporal data points and enable a direct comparison of the two sides.

A simulation parameter can be selected and is mapped via a colormap shared in (A) and (D).
Technically, the framework is capable of showing arbitrary scalar fields.
However, in this study, we focused on velocity magnitude and pressure as these are the conventional attributes utilized in clinical routine.
We use the Viridis colormap, as its colors are perceptually uniform and can also be perceived by most forms of color blindness~\cite{liu2018somewhere}.

\begin{figure}[tb]
    \centering
    \includegraphics[width=0.7\linewidth]{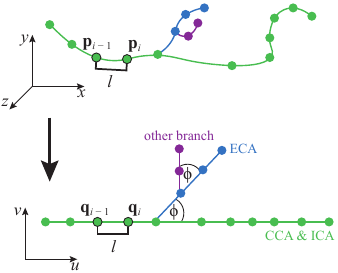}
    \caption{The map creation follows this blueprint. The branches are converted into lines, keeping the length $l$ between the segments.}
    \label{fig:flattening_schematic}
\end{figure}
\subsection{Domain Overview}
We show an initial overview of the 3D geometry as a surface rendering to give a general impression of the morphology.
The colormap encodes parameter values on the surface.
This view can also be set to display a streamline rendering (see Figure~\ref{fig:overview} (A)), showing the structure of the flow and the value distribution inside of the vessel for the selected parameter.
We chose integral lines instead of glyphs or volume rendering, as they are best suited to reveal occurring turbulence.
We found that physicians tend to evaluate singular time steps, e.g., at peak systole where the highest flow occurs.
Therefore, we chose to display streamlines rather than other types of integral lines or animated particles because they display the state of the flow field at exactly one time-step.
This aids in the overview task (\textbf{T1}) and minimizes necessary interaction.

The streamline seeds are placed in 100 random cells of each carotid volume mesh, resulting in a display that broadly covers the domain.
We use the adaptive Runge-Kutta integration scheme~\cite{kutta1901beitrag, runge1895numerische} of 5th order with a maximum propagation of $0.2$ meters (about the length of the carotid).
When rendering the streamlines we apply two techniques that have proven to be beneficial in similar scenarios.
First, we use front-face culling to add context information about the vessel wall without occluding the lines.
Second, we render implied tubes instead of lines, which are shaded to enhance depth perception.

\begin{figure}[tbh]
    \centering
    \includegraphics[width=1.0\linewidth]{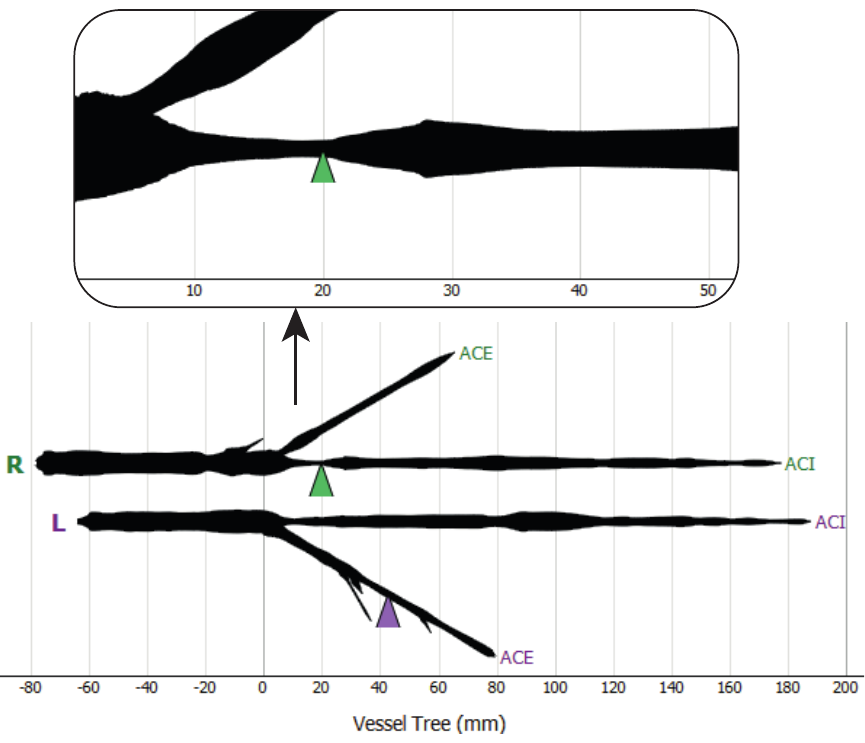}
    \caption{Map of the carotids used for overview and navigation. The line thickness encodes the vessel radius. Zooming in shows that the stenoses is about 20 millimeters long.}
    \label{fig:map}
\end{figure}
\subsection{Vessel Map}
In 3D depictions of vessel trees, self-occlusions are usually unavoidable.
Considerable user interaction is required to assess all parts of the geometry.
This makes the diameter of the vessel lumen difficult to compare across the full domain and means that developing stenoses might be overlooked.
In the past years, dimensional reduction techniques were effective to overcome these types of issues~\cite{kreiser2018survey, lichtenberg2020parameterization}.
In this sense, we construct a 2D map of the segmented vasculature to enable at-a-glance assessments of the diameter and increase comparability of data sets based on this abstract representation.

In the map the CCA and ICA are fixed to a straight line, providing an overview of this clinically relevant segment.
Neighboring vessels, like the ACE, are drawn splitting off at a fixed angle, e.g., $\phi = 30^\circ$, to indicate branch locations and give the map a tree-like appearance.
Computation of the map is performed as follows:
We treat each branch of the vessel centerline as a curve described by 3D points $\mathbf p_1, ..., \mathbf p_n$ connected in ascending order.
This curve is straightened to a line made up of corresponding 2D points $\mathbf q_1, ..., \mathbf q_n$.
The main branch (CCA and ICA) is laid out first and side branches are recursively connected at their relative positions.
Given the recursion depth $d \in \mathbb N_0$ and a start point $\mathbf q_1$ (which is either at $\mathbf 0$ for $d = 0$ or the branch location otherwise), the mapping for one curve is defined by:
\begin{equation}
\mathbf q_i = \mathbf q_{i-1} + \begin{pmatrix} \cos (\phi \cdot d) \\ \sin (\phi \cdot d) \end{pmatrix} \cdot ||\mathbf p_i - \mathbf p_{i-1}|| \quad \forall i \in [2,n]
\end{equation}
This procedure preserves the geodesic distance between points, showing the accurate length of each vessel segment in the map.
A schematic of this approach is shown in Figure~\ref{fig:flattening_schematic}.

As mentioned earlier, we already retrieved the radius information of the inscribed sphere at each sample location of the centerline.
We now map this information to the flattened representation by scaling the thickness of the rendered line.
We first used a linear scaling, however, we found that differences in the line thickness did not become obvious.
We overcame this problem by using a non-linear scaling which shrinks ``small'' radii and enlarges ``large'' ones.
This emphasizes narrow regions and thus highlights possible stenoses.
Small and large radii are distinguished with a threshold value $t$.
We use $t = 3$ millimeters, which is the average radius of the carotid in a healthy adult.
Each radius $r$ is then scaled by
\begin{equation}
    s(r) = \frac{t}{t^a} r^a,
\end{equation}
where $a \in \mathbb R \geq 1$ determines the ``linearity'' of the scaling.
A value of $a = 1$ produces a linear scaling.
We found stenoses to be suitably emphasized at around $a = 2$.

The vessel map is constructed for each of the two carotids.
Then, the maps are displayed in a combined depiction, where we flip the left-side carotid vertically and align the vessel trees in a stacked format (see Figure~\ref{fig:map}).
We register them horizontally at the bifurcation, as this is the defining landmark.
Our plot allows zooming and panning and we label the horizontal axis with millimeters.
This way, the flat depiction can be used to measure the length of a stenosis in the ICA.
Large branches are automatically labeled.

\begin{figure}[tb]
    \centering
    \includegraphics[width=1.0\linewidth]{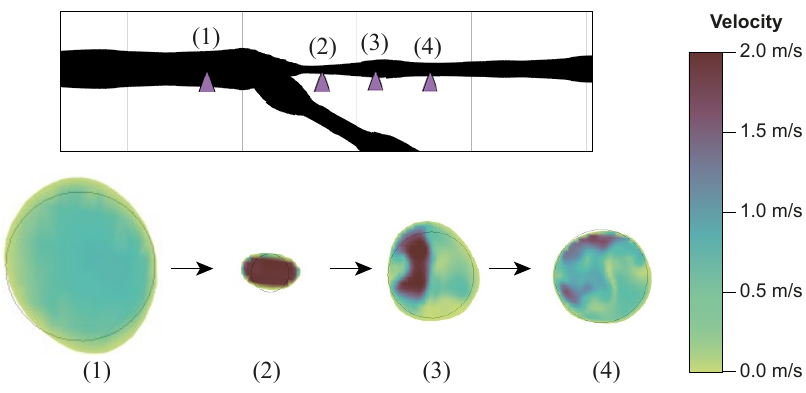}
    \caption{Dragging the probe reveals the hemodynamics of stenoses. Usually the velocity profile should be homogeneous with slower flow in proximity to vessel walls (1). Velocity increases in stenoses (2), followed by visible turbulence (3 and 4).}
    \label{fig:probing}
\end{figure}
\subsection{Flow Rate Graph}
To give the user a sense of the temporal dimension, we plot the inflow volume of each side over time, resulting in a view showing superposed line graphs (see Figure~\ref{fig:overview} (B)).
This is a beneficial depiction for two reasons:
First, temporal patterns regarding the vascular system are commonly based on the cardiac cycle.
The flow rate graphs provide an overview of the simulated cardiac cycles and, therefore, tie in with the background knowledge of the users.
Second, the volumetric flow rate is a vital global parameter, which essentially determines the quantity of oxygen a vessel can transport.
As the flow rate cannot be easily measured in-vivo, it is typically not used in practice.
In a simulation, however, the in- and outflow volumes are always known or can be derived.
Therefore, they should be visually integrated to provide an impression of the transported blood volume.

As two graphs are used, the flow rate of the left and right carotid can be directly compared.
Further, the display is leveraged to show the flow rate inside specific vessel branches.
A second line is rendered per side in a dashed style, illustrating the flow in one additional branch.
This can be used to compare how much of the total flow is received by one branch, e.g., the ICA.
Displaying the flow rate in more than one additional branch per side would clutter the superposed line graphs.
Therefore, we select this branch based on where the spatial probe is located.
As the simulation assumes incompressibility of the fluid, the flow rate is equal at every location in one vessel branch.

\begin{figure}[tb]
    \centering
    \includegraphics[width=1.0\linewidth]{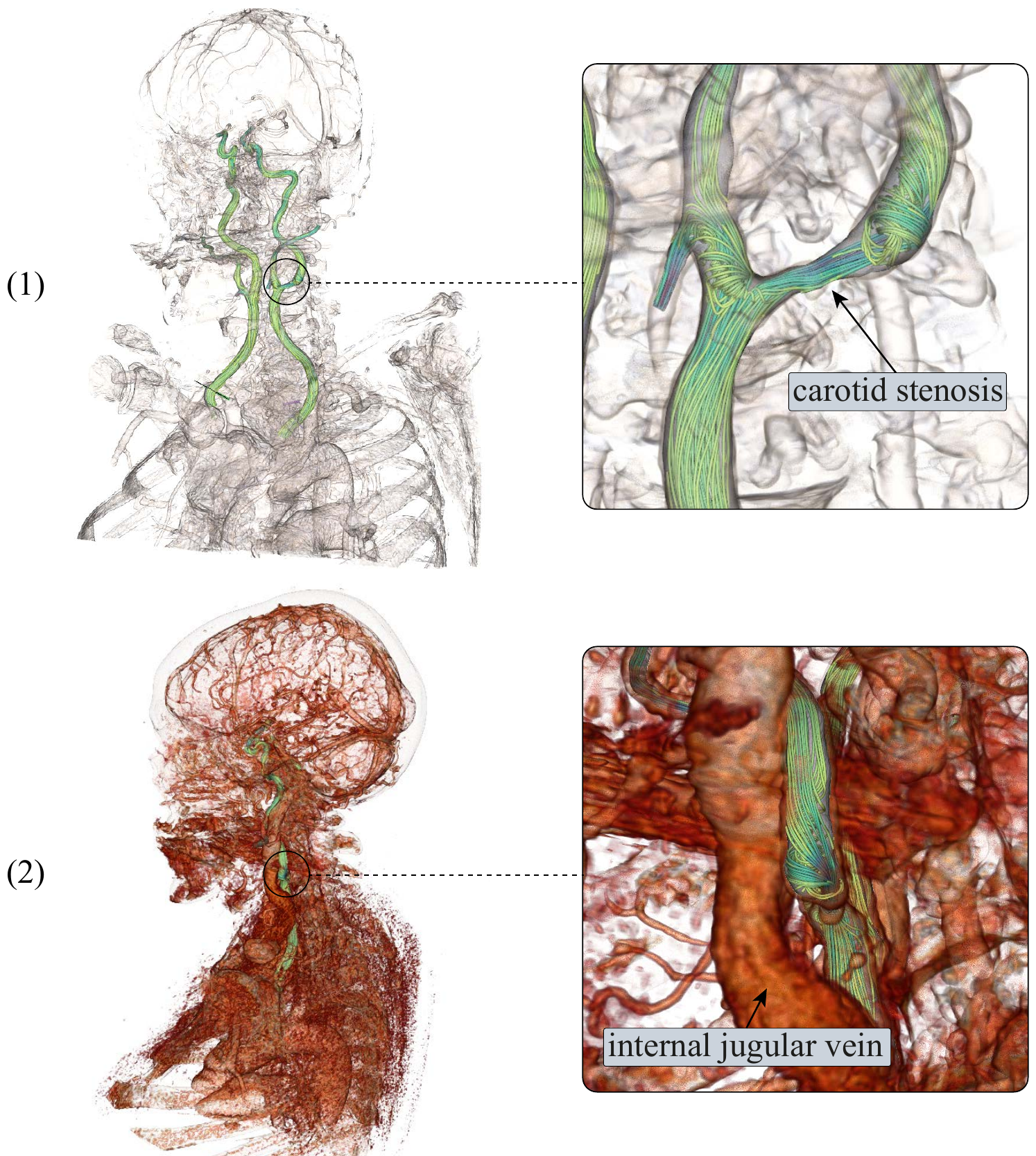}
    \caption{The contextual volume rendering shows that a stenosis is located in the upper neck area (1). Adjusting the thresholding and opacity (2) reveals that the internal jugular vein is in front of this stenosis, potentially complicating surgery.}
    \label{fig:DVR}
\end{figure}
\subsection{Navigation and Probing}
We aim to visually guide users to derive relevant insights quickly, thereby minimizing the necessary interaction.
Simultaneously, we want to give profound access to the spatio-temporal data space (\textbf{T2}), allowing all parameters to be thoroughly assessed.
In our use case, the temporal dimension appears to be utilized less during exploration.
It is often set at the beginning of the analysis process and is only infrequently adjusted.
Therefore, we approach this optimization problem by partitioning the controls for data exploration into temporal and spatial navigation.

We use the flow rate graph as a tool for temporal navigation.
It reflects the cardiac cycle, providing an intuitive sense of time.
Points of interest, typically the systolic and diastolic time-steps, can be easily discerned.
We integrate a slider that can be dragged over the plot, which selects the nearest time-step in the data series and updates all views accordingly.

Probes are placed via the vessel map, as it presents the full spatial domain in one depiction and is designed to highlight relevant regions, where stenoses could be located.
We use a triangular marker to indicate the probe position, as it is conveniently selectable and does not hide information about the vessel thickness.
This omits the need for ambiguous placement interactions in the 3D domain.
There exists a variety of options for how a spatial probe could be implemented.
In the discussions with the domain experts, we concluded that vessel cross-sections, i.e., plane cuts, would be most useful.
They show the vessel shape for morphological assessments and the parameter distribution (e.g., the velocity profile) for hemodynamic analysis.
%
%
The optimal plane cut is usually orthogonal to the vessel direction.
This is the most accurate representation of the diameter, which would otherwise be elongated in at least one axis.
Therefore, we compute the direction of the cutting plane using the centerline vectors as normals.
Ultimately, this means only minimal interaction is required to place the plane and none to orient it.

The resulting cut could be shown in the 3D view.
However, such a display of the vessel cross-section would suffer from perspective distortion and the viewer's direction would need to be changed depending on the probe position, to see it fully.
Therefore, we implemented a separate 2D view of the cross-section, which is aligned on the vessel center (see Figure~\ref{fig:overview} (D)).
We overlay the probe view with a millimeter grid to enable measurements and size estimations.
Simultaneously, we indicate the probe location in the 3D view with the same $20 \times 20$ millimeter grid.
Furthermore, the computed diameter is displayed below the probe view, as it is an important quantifiable parameter.
It is also hinted at as a circle to allow validation of the numerical output in regions where the vessel is more elliptical.
For direct comparisons (\textbf{T3}) we display two probe views, one for the left and right carotid.
We differentiate between the two sides in all views using labels and unified colors.
To facilitate a fluent evaluation of the parameter domain along the vessel centerline, the probe can be interactively dragged over the vessel map.
The probe view then shows an animation of how the diameter changes and helps to analyze how turbulences behave (see Figure~\ref{fig:probing}).

\begin{figure}[tb]
    \centering
    \includegraphics[width=1.0\linewidth]{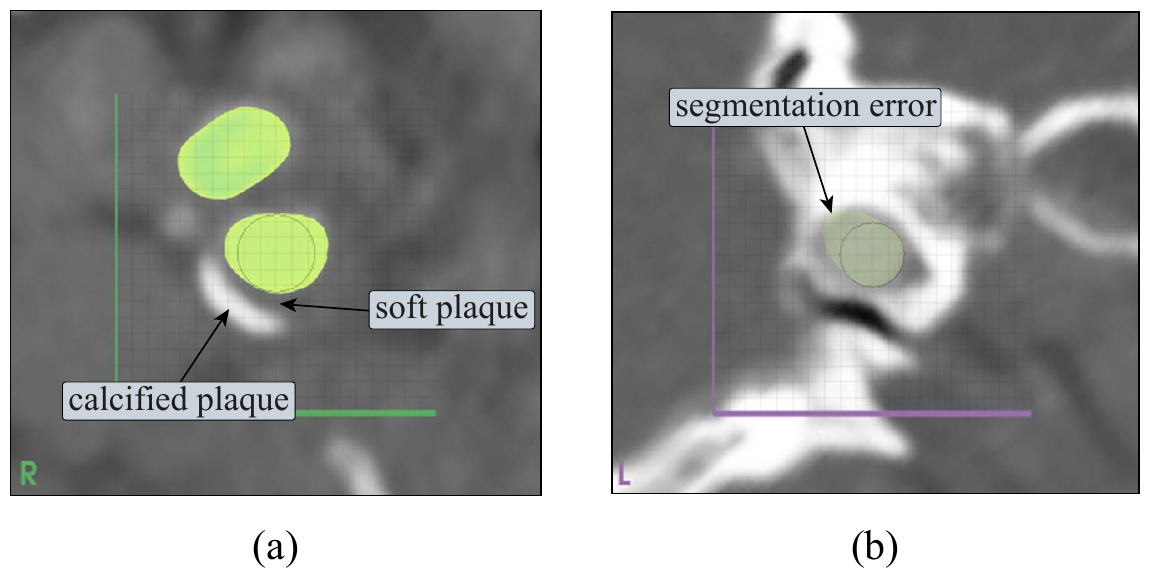}
    \caption{The probe can also sample the CTA volume. This allows to assess plaque (a) and validate the segmentation at crucial locations (b).}
    \label{fig:context2D}
\end{figure}
\begin{figure}[tb]
    \centering
    \includegraphics[width=1.0\linewidth]{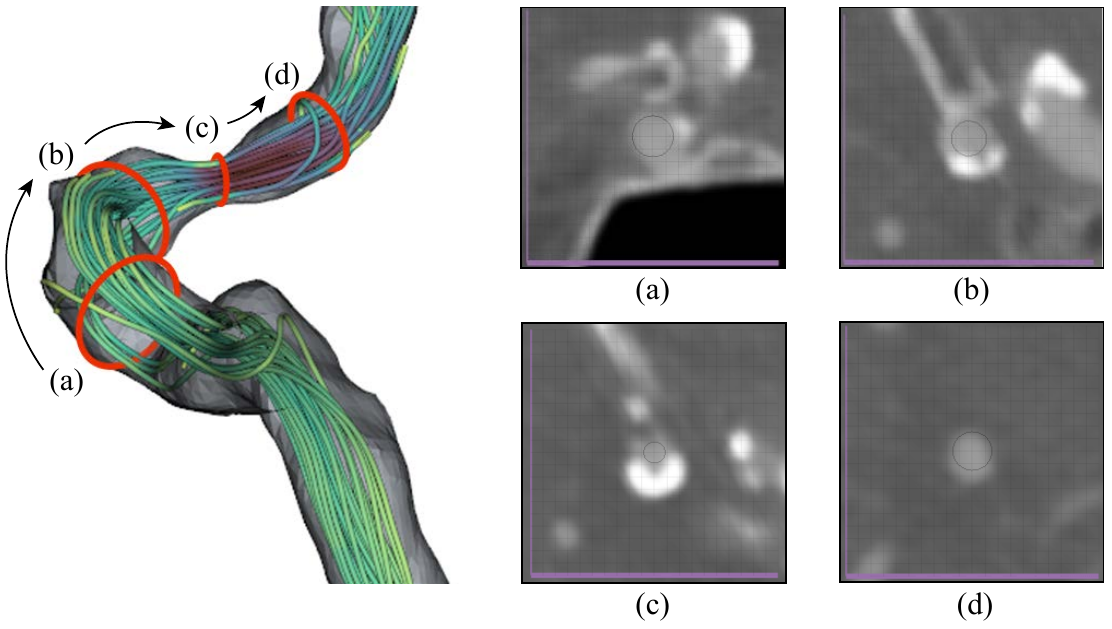}
    \caption{The probe follows the vessel direction when dragged. This gives an accurate depiction of the lumen diameter, which is essential when assessing the stenosis at (c). Probing the CTA volume unveils the cause of this stenosis $-$ a buildup of plaque on the vessel wall.}
    \label{fig:context_drag}
\end{figure}
%
\subsection{Contextualization}
We determined that, for a full assessment, surrounding structures need to be evaluated (\textbf{T4}).
To give a sense of the global context, we embed the original CTA data on-demand into the 3D scene using volume rendering.
The definition of transfer functions for volume rendering is a demanding task.
As the depiction is only intended to show additional context information, we simplify this for the user.
We use preset transfer functions (intensity to color, intensity to opacity, and gradient to opacity), which we derived from functions that have been found suitable in coronary artery visualization~\cite{glasser2010automatic}.

Some control, however, is given to the user, as the region of interest can vary depending on the task (e.g., whether vessels or tissue should be shown).
Also, the value ranges of CTAs are not universal due to differences in the contrast agent diffusion.
Two sliders are implemented for this purpose:
The first shifts the window of the visible value range and the second defines the overall opacity.
This way, the user can quickly adjust the density region shown and determine how dominant the volume rendering should be.
The final scene is rendered using volume ray casting with composite blending.
This method allows a hybrid depiction of the polygonal models (segmented carotids and streamlines) within the CTA volume.
Figure~\ref{fig:DVR} illustrates how this could facilitate far context analysis.

The contextualization is also extended to the 2D probe view, to show the near context of tissue and possible plaque surrounding the vessel cross-section.
We sample the CTA volume with the probe plane.
The resulting texture is optionally displayed in the background of the probe view.
It is rendered with shades of gray similar to the standard CTA representations.
We map the Hounsfield Unit range $[-800, 1200]$ linearly to the gray levels $[0, 255]$.

The primary purpose of the contextual CTA slice is to enable the analysis of plaque on the vessel walls.
However, it is also useful to validate the segmentation, as demonstrated in Figure~\ref{fig:context2D}.
The user can adjust the opacity of the model cross-section to reveal all of the original data if desired.
With this representation, the CTA volume can be explored similarly to the spatial simulation domain.
Dragging the probe across the vessel map reveals the varying shape of the carotid, occurrences of plaque, and changes in the contrast agent diffusion.
Aligning the view plane orientation to the vessel direction allows a consecutive evaluation of the vessel following the orthogonal cross-section (see Figure~\ref{fig:context_drag}).
This gives a more accurate view of the vessel morphology as compared to using a fixed slice orientation (the typical exploration method in practice).
Furthermore, no deformation of the volume is introduced, as is the case when applying planar reformation methods~\cite{wesarg2006localizing}.
\section{Evaluation}
We conducted qualitative reviews with seven domain specialists (2 female, 5 male; 26-61 years old), including the three experts we spoke to during the stakeholder analysis. 
To evaluate if the methods we adapted aid in the defined tasks, we first individually interviewed two radiologists, one neurologist, and one neuroradiologist who routinely treat stroke patients.
The neurologist is also co-author of this paper but had not used the developed application before.
We reviewed the concept of flow simulation based on CFD and explained the pipeline used to generate our data sets.
Then, we introduced them to our framework and demonstrated how it can be used to simultaneously explore hemodynamic and morphological information.
We explained each view and the ideas behind it.
Then, we consecutively presented three data sets of differing types (full stenosis, developing stenosis, and no stenosis).
For each set, we asked the participants to interact with the framework themselves and explore the data as if they were to perform a diagnosis.
We inquired what they could tell us about the data sets and which information they could gain from the visualizations.
We encouraged them to think-aloud during this process.
Afterward, participants filled out a questionnaire.
We associated each view with a series of statements relating to the tasks the view was designed for.
Participants were to rate how much they agreed with the statements on a five-point Likert scale ($--$, $-$, $\circ$, $+$, $++$).
We also showed two pictures of the vessel map, one with linearly scaled diameter mapping and one based on our non-linear scaling method.
Without knowing the scaling parameters, participants were to choose on which map they could better identify possible stenoses.

While physicians can evaluate how useful the proposed methods would be in practice, they are less aware of which exploration and visualization approaches exist to display CFD data.
To determine whether we chose and transferred fitting flow visualization techniques and which other options might exist, we extended the evaluation.
We interviewed two researchers working with CFD simulations of the carotids and a flow visualization expert with a research background in aneurysm hemodynamics.
We followed a similar evaluation procedure, however, we did not ask the experts to perform mock-up diagnoses.
Instead, we demonstrated the steps clinicians took when exploring the data sets, summarizing what was important during clinical assessments.
In the questionnaire, we omitted the parts relevant only to physicians, e.g., regarding the analysis of plaque in embedded CTA data.
In general, we steered the discussions towards whether other visualization methods were known and could be applied.
\section{Results and Discussion}
All participants quickly understood the linked views and were able to explore the data without noticeable difficulties.
When the clinicians examined the example data sets, they mostly defaulted to a single systolic time-step and analyzed the domain at this point of peak flow.
They immediately identified the stenoses (\textbf{T1}) and predominantly used the framework to probe the surrounding velocity magnitudes and vessel diameters.
For this task, the vessel map was extensively used as a navigational tool.
All experts preferred the non-linear over the linear diameter scaling.
They did not see an issue with over- or underestimation of the diameter, as the map was mainly used for localization of potential stenoses, which were then assessed in detail using the probe view.
%
%
The radiologists pointed out that dragging the plane probe along the carotid shows cross-sections well suited for probing (\textbf{T2}), as they could instantly evaluate the lumen diameter, shape, and the condition of the wall over a vessel segment.
In some cases, participants placed a probe at an equivalent point on the opposite carotid to compare the diameters (\textbf{T3}).
All physicians extensively used the contextual CTA slice around the probe (\textbf{T4}) and appreciated the integration of hemodynamic information on top of the CTA.
However, participants also mentioned that they felt uncertain about the correctness of the computed values.
The neurologist made us aware of an issue we did not consider:
Parts of the carotid branches in the cranium form a vascular short circuit between the left and right sides.
Due to this connection, a severe stenosis in one side often leads to a reversed flow in the following section of the ACI.
As we currently simulate flow in both carotids separately, we cannot represent this phenomenon.
Therefore, we argue that our approach should be improved by taking further cranial branches into account.
\begin{figure}[tbh]
    \centering
    \includegraphics[width=1.0\linewidth]{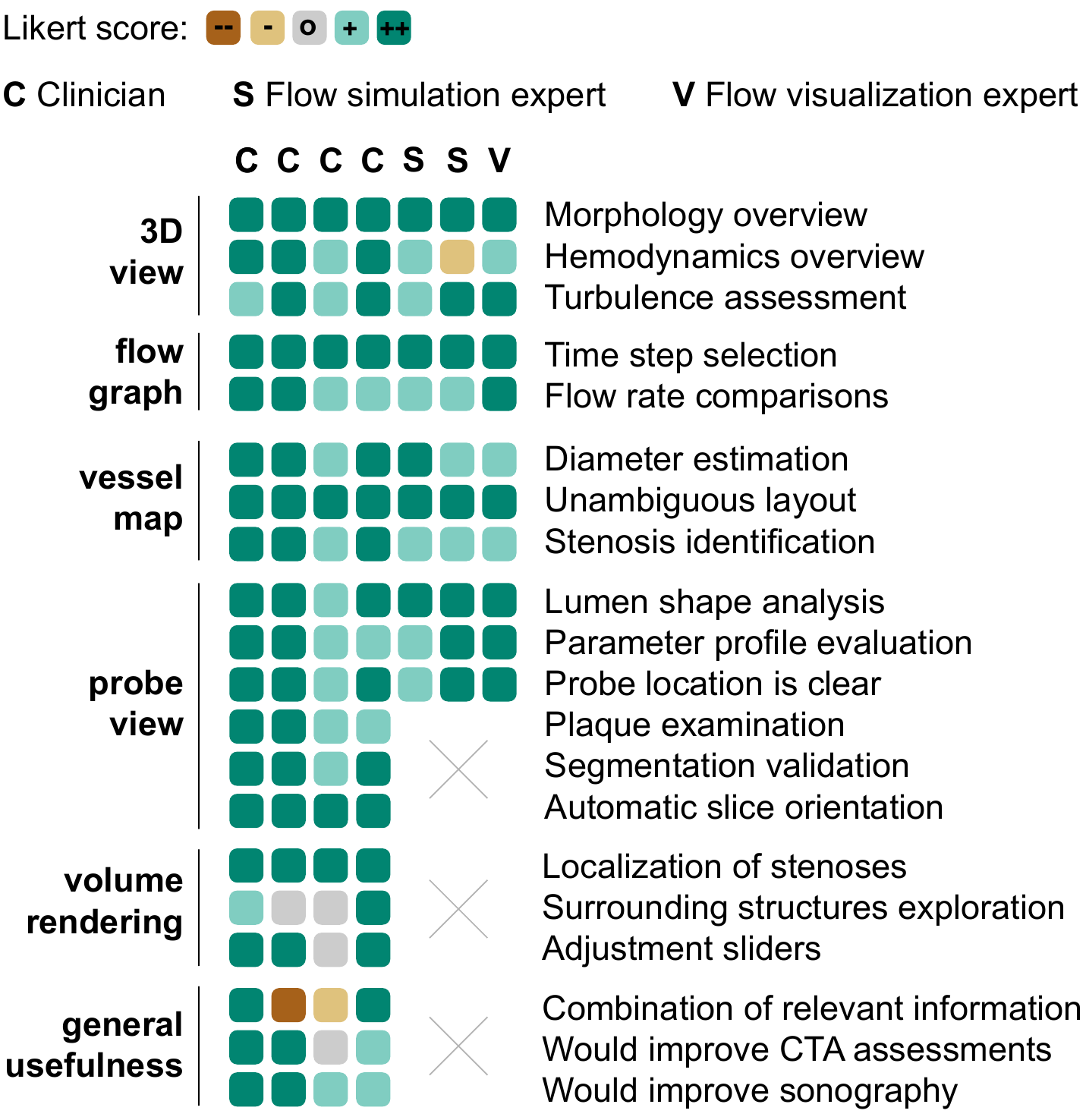}
    \caption{Results of the questionnaire with color-encoded Likert scores. Each box shows the answer of one participant.}
    \label{fig:evaluation_results}
\end{figure}

The interviewed simulation and visualization experts generally found the selected methods suitably chosen and adapted.
They had only minor recommendations and improvement suggestions.
One CFD expert proposed to include quantified information on stenoses, e.g., the minimum diameter at a predicted stenosis.
This could be included in the vessel map view for a faster initial comparison of multiple stenoses.
The flow visualization expert suggested to extract plaque geometry in addition to the vessel lumen and integrate it into an overview depiction (the 3D or the map view).
Calcified plaque could be automatically segmented from CTA and used to indicate regions where pathological processes occur.

The questionnaire results can be seen in Figure~\ref{fig:evaluation_results}.
They reflect the generally positive impressions of all participants. 
Still, some aspects should be discussed.
The volume rendering was well received for the localization of stenoses, but was less used to explore surrounding structures, as participants found the noise distracting.
Filtering could be applied to remedy this issue, however, this would not be a trivial task, as context information should not be occluded.
Different types of structures (e.g., tissue, vessels, or bones) can be of interest, which makes the definition of automatic filters difficult.

One assertion in the survey led to disagreements: \textit{The framework appropriately combines hemodynamic (as in duplex sonography) and morphological information (as in CTA).}
Two participants argued that the hemodynamic data shown is simulated not measured, i.e., the framework does not combine information but rather shows additional data.
One could suggest that we should have phrased the statement differently, to better reflect our actual workflow.
Then, we could argue that the validity will increase with better simulations available.
As this would likely not affect how the data should be visualized, we could dismiss the issue as outside of our research domain.
However, we want to draw attention to this point, as it is, again, indicative of an important challenge:
Many clinicians see the validity of the simulation results as a barrier to the practical adoption of CFD methods.
We revisited this discussion with researchers from both sides (medical and CFD) and singled out a core problem: advances in the CFD domain are rarely brought to the attention of physicians, and the requirements of clinicians are often unknown to researchers in the field of CFD.
This bottlenecks clinical integration of computed flow data, which would be necessary for large trials that could support data validation.
We believe that visualizations of simulated data could be key to solving this issue, as they can provide clinicians with appropriate views on the computation results, which has high potential to improve interdisciplinary communication.
Furthermore, our discussions led to two options for how visualizations could directly address the uncertainty around simulated flow data:
A possible approach would be to register measured data from CTA and sonography in a combined visual display.
Effectively, this would either require to integrate the sonography data at the end of the pipeline or to replace the computed flow fields completely.
This is a challenging objective due to the low resolution and high noise associated with ultrasound assessments.
Also, flow information in cranial regions can often not be obtained this way.
Another option would be to use sonographic measurements to initialize boundary conditions for the simulation in areas where sonography can be performed.
Ideally, this would combine the advantages of sonography (high accuracy) and CFD (high resolution).
\section{Conclusion and Future Work}
In this design study, we investigated how medical flow visualization can be applied to enhance the diagnostic evaluation of carotid stenoses.
We developed and tested an application with four linked views, which enables physicians to explore the vessel morphology in combination with simulated hemodynamics.
Our findings indicate that many concepts from in flow visualization research can be transferred to match specific tasks in this scenario.
We found geometry abstractions to support overview tasks and the identification of points of interest.
We stress that this appeared particularly useful for spatial navigation, as demonstrated with the flattened vessel tree we employ to place data probes.
Further, we would like to emphasize the importance of contextualization in medical visualization, which is a key feature of this work.
Context information can not only be used to localize stenoses but also to explore important additional features (such as plaque), and perform on-the-fly validations.
We also realized that some caveats should be considered when developing visualizations with respect to stenosis analysis.
For clinical use cases, the local view on flow patterns and parameters in the vicinity of stenoses should be complemented by a more global perspective.
To gain a full picture, it is crucial that the general distribution of blood flow and the occurrences of multiple stenoses can be explored.
Furthermore, we found the temporal dimension to be of less importance than we first anticipated.
From our evaluations and discussions, we assume that a depiction focused on the systolic and diastolic values could be more useful in practice than the integration of a high number of time-steps.
Lastly, it should be noted that CFD can generate an extensive amount of data.
The exploration of this data in application scenarios, such as medical diagnostics, can be cumbersome and prevent the integration into routine.
It is therefore essential that the core tasks of users are distilled first, followed by tailoring methods to match these tasks exactly, aiming to minimize necessary interaction.
In this study, developing the right exploration techniques played a similarly important role to choosing and adapting visualization methods.

In the future, a central objective should be to increase the validity and, therefore, the applicability of hemodynamics computed with CFD.
We mentioned two options for how visualization could advance these efforts.
First, different modalities (e.g., CTA and sonography) could be visually integrated.
%
%
Second, a larger vascular domain could be segmented to compute the correct flow properties of arterial short-circuits.
This would pose the question of how such an extended domain could be depicted without cluttering the view.
Furthermore, we prioritized parameters that are widely used, such as flow velocity and vessel diameter.
However, CFD is often employed to generate complementary attribute fields, which could also play a role in pathogenesis.
To foster a better understanding of such additional factors, future work should aid in the investigation of occurring patterns.
%
%
Cohort visualizations could facilitate this endeavor by enabling comparisons across patient collectives.
Potentially, this would allow further insights to be derived which could then be applied to supplement diagnostic assessments.

\section*{Acknowledgements}
This projected was partially funded by the Carl Zeiss Foundation and the BMBF Joint Project 05M20SJA-MLgSA.

\bibliographystyle{eg-alpha-doi}  
\bibliography{bibliography.bib}        


\end{document}